\documentclass{ws-ijmpe}
\usepackage{pst-all}

\begin{document}

\markboth{Sprenger, Nicolini, Bleicher}{Quantum Gravity signals in neutrino oscillations.}

\catchline{}{}{}{}{}
%%%%%%%%%%%%%%%%%%%%%%%%%%%%%%%%%%%%%%%%%%%%%%%%%%%%%%%%%%%%%%%%%%%%

%\title{NEUTRINO OSCILLATIONS AS A NOVEL PROBE FOR A MINIMAL LENGTH}
\title{QUANTUM GRAVITY SIGNALS IN NEUTRINO OSCILLATIONS}

\author{\footnotesize MARTIN SPRENGER\footnote{sprenger@fias.uni-frankfurt.de}}

\author{\footnotesize PIERO NICOLINI\footnote{nicolini@th.physik.uni-frankfurt.de}}

\author{\footnotesize MARCUS BLEICHER\footnote{bleicher@th.physik.uni-frankfurt.de}}

\address{Institut f\"ur Theoretische Physik \& Frankfurt Institute for Advanced Studies (FIAS)\\
Johann Wolfgang Goethe-Universit\"at, 60438 Frankfurt am Main, Germany\\
}

\maketitle

\begin{history}
\received{(received date)}
\revised{(revised date)}
%\accepted{(Day Month Year)}
%\comby{(xxxxxxxxxx)}
\end{history}

\begin{abstract}
We investigate the effect of a Quantum Gravity-induced minimal length on neutrino oscillations.
The minimal length is implemented in a phenomenological framework, allowing us to make predictions independently of any fundamental approach.
We obtain clear minimal length signatures and discuss their observability in current and future experiments.
We present an overview over other scenarios in which the minimal length leaves its signature and show new results concerning minimal length thermodynamics.
\end{abstract}

\section{Introduction}

One of the central questions every theory of Quantum Gravity has to settle is the nature of space-time at the fundamental scale.
Candidate theories so far do not agree on whether space-time becomes discrete, stays continuous or both \cite{Kempf2010}.
However, it is widely believed that space-time cannot be probed to arbitrarily small distances in Quantum Gravity.
The following heuristic argument supports this idea:
Consider a particle that is used to probe space-time.
To probe smaller structures of space-time, the energy of the probe has to increase.
At one point, the energy of the probe is so large that it collapses into a black hole and all information is lost in the event horizon.
This happens when the region over which the particle is localised, the Compton wavelength, and the Schwarzschild radius of the probe are of the same order, i.e.
\begin{equation}\lambda_{C}\approx r_S\label{eq:epl}\end{equation}
Structures smaller than this scale $\ell^{-1}$ cannot be resolved and lose operational meaning.
A maximum resolution or, in turn, a minimal length emerges.
From Eq.(\ref{eq:epl}) one usually expects $\ell^{-1}$ to be of the order of the Planck energy.
However, only $\ell^{-1}\gtrsim 1~\mathrm{TeV}$ is established experimentally \cite{Kapner2007}.
Due to the discrepancy of the above values, we will use $\ell^{-1}$ as a free parameter in this contribution and try to constrain its value by studying the modifications due to the minimal length in neutrino experiments.
We will implement the minimal length in a phenomenological approach which allows us to make general predictions, independent of any particular fundamental theory.
Neutrinos are ideally suited for minimal length studies as they are only weakly interacting and can propagate freely for very long distances.
A minimal length effect might therefore pile up beyond detectable thresholds.\\
This contribution is organised as follows:
In section \ref{sec:Model} we will implement a minimal length in a phenomenological framework.
We present the idea of neutrino oscillations in section \ref{sec:osci}, before studying the modifications due to the minimal length in section \ref{sec:osci_ml}.
To conclude, we provide a brief outlook on other studies in section \ref{sec:outlook}.

\section{Model}
\label{sec:Model}
\begin{figure}
\centering
\includegraphics[scale=0.6]{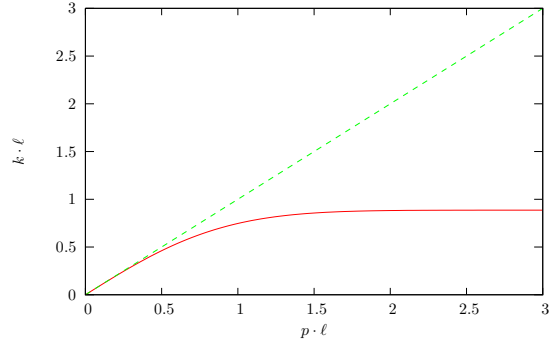}
\caption{Modified dispersion relation in a GUP theory (red solid line) and in the classical case (green dashed line). The modified dispersion relation is linear for small $p$, but saturates for high momenta.}
\label{fig:disp_rel}
\end{figure}

To implement a minimal length, we use inspiration from String Theory.
In String Theory, a minimal length emerges by a modification of the uncertainty principle
\begin{equation}
\Delta x\Delta p\geq \left(1+c\frac{\left(\Delta p\right)^2}{M_f^2}\right),
\label{eq:st_gup}
\end{equation}
where c is a constant and $M_f\sim\ell^{-1}$ is the fundamental scale, see \cite{Konishi1990}.
This inequality cannot be satisfied for arbitrarily small $\Delta x$, as is easily checked.
Therefore, Eq.(\ref{eq:st_gup}) indeed gives rise to a minimal length.
Recalling the connection between the uncertainty relation of observables and the non-commutativity of the corresponding operators in quantum mechanics, we can implement the minimal length by modifying the operator algebra.
The most general algebra that gives rise to a minimal length reads
\begin{align*}
	\left[\hat{x}_i,\hat{p}_j\right]&=i\delta_{ij}\left(1+f\left(\hat{p}^{\, 2}\right)\right), \\
	\left[\hat{p}_i,\hat{p}_j\right]&=0, \\
	\left[\hat{x}_i,\hat{x}_j\right]&=-2i\hat{L}_{ij}f'\left(\hat{p}^{\,2}\right),
\end{align*}
where $\hat{L}_{ij}$ is an angular momentum operator.
The choice $f\left(\hat{p}^{\, 2}\right)=\beta \hat{p}^{\, 2}$ leads to a commutator of the form of Eq.(\ref{eq:st_gup}), while in theories inspired by noncommutative geometry\cite{Smailagic:2003rp} one chooses $f\left(\hat{p}^{\, 2}\right)=e^{\ell^2\hat{p}^{\, 2}}-1$.
However, both choices lead to the same physical effects.\\ 
For simplicity, we keep momenta commuting.
The position commutator is then fixed by the Jacobi identity.
As momenta are commuting, we still have a momentum eigenbasis.
This is a very convenient basis to work with, as all changes due to the minimal length can be accounted for by modifying the momentum space integration measure as
\begin{equation}
d^3p\rightarrow \frac{d^3p}{1+f\left(\hat{p}^{\, 2}\right)}.
\end{equation}
Since we have a continuous momentum eigenbasis, momenta are unbounded in GUP theories, as explained in detail in \cite{Hossenfelder2006}.
On the other hand, it is clear that for all physical states the wavelength of a particle has to be larger than the minimal length which in turn implies that the wave vector is bounded from above in GUP theories.
Combining these two ideas, we see that the dispersion relation $k(p)$ will be non-linear.
Generalising to four-vectors, $\omega(E)$ will be of the same functional form as $k(p)$.
For the noncommutative geometry-inspired theories the modified dispersion relation is given by
\begin{equation}
\omega(E)=\frac{\sqrt{\pi}}{2\ell}\mathrm{Erf}\left(\ell E\right)
\label{eq:mod_disp_rel}
\end{equation}
with $\mathrm{Erf}(x)$ being the error function, which is plotted in figure \ref{fig:disp_rel}.
As is visible, the dispersion relation is linear for small values of $E$ but saturates for large $E$.
For more details on GUP theories see \cite{Kempf1995}.

\section{Neutrino Oscillations}
\label{sec:osci}
Before discussing the effects of the minimal length on neutrino oscillations, let us briefly recall the basic ideas of conventional neutrino oscillations.
In several experiments, it was established that neutrinos can change their flavour during free propagation (for a review see \cite{Winter2010}).
The most common and successful interpretation of this effect is that neutrinos have masses and propagate freely in mass eigenstates rather than in flavour eigenstates.
For three flavours, these oscillations can be parametrised by 6 parameters: 2 squared mass differences $\Delta m^2_{ij}$, 3 mixing angles $\theta_{ij}$ and a CP-violating phase $\delta_{CP}$.
The transition probability between flavour eigenstates $\alpha$ and $\beta$ is then given by
\begin{equation}
P(\nu_\alpha\rightarrow\nu_\beta)=\sum\limits_{k,j=1}^3 U_{\alpha k}^\ast U_{\beta k} U_{\alpha j} U_{\beta j}^\ast e^{-i\frac{\Delta m_{kj}^2}{2E}L},
\label{eq:cl_prob}
\end{equation}
where $L$ is the propagation length of the neutrino and $E$ its energy.
$U$ is the Pontecorvo matrix describing the basis change from flavour eigenbasis to mass eigenbasis and is parametrised by $\theta_{ij}$ and $\delta_{CP}$.
The exponential in Eq.(\ref{eq:cl_prob}) gives rise to the oscillatory behaviour.

It should be noted that if the propagation length $L$ becomes too large with respect to the oscillation length
\begin{equation}
L_O=\frac{2E}{\Delta m^2_{kj}},
\end{equation}
the oscillations become very rapid and decohere.
Therefore, to obtain a clear signal one has to take care that the propagation length is of the order of the oscillation length.

\section{Neutrino Oscillations with a Minimal Length}
\label{sec:osci_ml}
\begin{figure}
\centering
\input{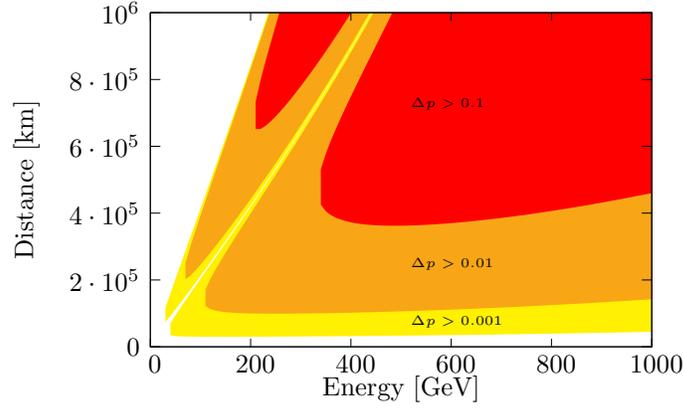}
\caption{Regions for neutrino energy and oscillation length for which a significant effect is observable with $\ell^{-1}=1~\mathrm{TeV}$.}
\label{fig:lands}
\end{figure}

In the derivation of Eq.(\ref{eq:cl_prob}) use is made of the dispersion relation $\omega=E$.
We can now implement the effect of a minimal length by using the modified dispersion relation Eq.(\ref{eq:mod_disp_rel}).
Going through the calculation of the transition probability from flavour eigenstate $\alpha$ to flavour $\beta$, we find the modified probability
\begin{equation}
P_{\ell}(\nu_\alpha\rightarrow\nu_\beta)=\sum\limits_{k,j=1}^3 U_{\alpha k}^\ast U_{\beta k} U_{\alpha j} U_{\beta j}^\ast e^{-i\frac{\Delta m_{kj}^2}{2E}\exp\left(-\ell^2E^2\right)L},
\label{eq:mod_trans}
\end{equation}
where $U$ is the same matrix as in the classical case.
%To get a rough estimate of the size of the minimal length effect, we apply the modified transition probability Eq.(\ref{eq:mod_trans}) to data from the MINOS experiment \cite{Adamson2011}, with the results shown in figure \ref{fig:minos}.
%\begin{figure}
%\centering
%\input{ms_poster}
%\caption{Modifications of the oscillation pattern for different values of $\ell^{-1}$.}
%\label{fig:minos}
%\end{figure}
%Clear deviations from the classical behaviour are only present for small values of $\ell^{-1}$, while the oscillation pattern converges quickly towards the classical curve for larger values of $\ell^{-1}$.
%From the data, we find a lower bound on the minimal length $\ell^{-1}\gtrsim 10~\mathrm{GeV}$, far below the best current bound $\ell^{-1}\gtrsim 1~\mathrm{TeV}$.
%However, increasing the neutrino flux by a factor of $100$ might push the bound into the $100~\mathrm{GeV}$ region.
%Therefore, the next generation of neutrino experiments will allow us to strengthen the bound, providing a new and original way to constrain the fundamental scale.
%
Looking for significant departure from classical behaviour, we study the quantity
\begin{equation}
\Delta p =\left|P(L,E)-P_\ell(L,E)\right|
\end{equation}
for a large range of energies and distances, with the results shown in figure \ref{fig:lands}.
We see that already for distances within the solar system, regions with $\Delta p>0.1$ exist for $\ell^{-1}=1~\mathrm{TeV}$.
Note that $\Delta p$ is an absolute difference of probabilities.
Therefore, $\Delta p>0.1$ corresponds to flux differences of $10\%$ or more, leaving a clear signature in the oscillation pattern.

It should be noted that the results in figure \ref{fig:lands} are robust against uncertainties in the oscillation parameters since we are looking at the difference between oscillation probabilities.
Another point that should be noted is that the oscillation length in the minimal length scenario is exponentially enhanced,
\begin{equation}
L_\ell=\frac{2E}{\Delta m^2_{kj}}e^{\ell^2E^2}.
\end{equation}
Therefore, neutrino oscillations are coherent over much larger distances for high energies.
In fact, for energies larger than the fundamental scale, oscillations are strongly suppressed.
High-energy neutrinos therefore do not oscillate.
This leads to a striking experimental signature, as high-energy neutrinos coming from galactic and extra-galactic point sources such as active galactic nuclei or gamma-ray bursts reach the Earth in their original flavour composition in the minimal length model and are not in a perfectly mixed state, as expected from standard oscillations.
These ideas can be tested in neutrino telescopes such as IceCube or ANTARES in near future.
For a derivation of Eq.(\ref{eq:mod_trans}) and more details see \cite{Sprenger2010}.

\section{Summary and Outlook}
\label{sec:outlook}
\begin{figure}
\centering
\includegraphics[scale=0.6]{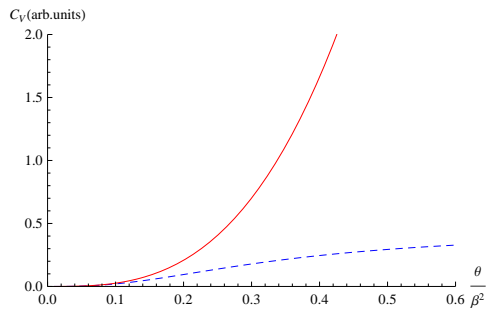}
\caption{Heat capacity of a photon gas with (blue dashed line) and without minimal length (red solid line).}
\label{fig:cv}
\end{figure}

In this contribution, we showed how minimal length effects can be modelled in a phenomenological approach.
The minimal length leaves clear signatures in neutrino oscillations.
With current statistics, these signatures are not observable in earthbound experiments but can lead to strong bounds on the minimal length for the next generation of neutrino experiments.
For solar-system distances the minimal length effect is enhanced and leads to flux differences of $10\%$ and more which would be easily observable.
Neutrino telescopes should be able to test the scenario proposed in this contribution in near future due to distinct signatures from the spectrum of high-energy neutrinos coming from galactic and extra-galactic point sources.
Other minimal length studies are in progress.
Minimal length effects could become interesting in thermodynamics as the effects might be enhanced due to the large number of particles.
In figure \ref{fig:cv} we show the heat capacity of a photon gas with a minimal length.

Instead of diverging, the heat capacity saturates, leading to a clear Quantum Gravity signature.
Other studies include black hole solutions with a minimal length.
These solutions are free of singularities and show a very different decay behaviour compared to classical black holes \cite{Nicolini2008}.
This might lead to observable signatures in the study of microscopic black holes at the LHC \cite{Nicolini2011}.
Other effects include a natural regularisation of the Casimir energy \cite{Garattini2010}, the prediction of a two-dimensional Planck scale spectral dimension \cite{Modesto:2009qc} and other novel scenarios in cosmology \cite{Mann:2011mm}, particle physics \cite{g-2} and black hole thermodynamics \cite{Nicolini:2010nb,Nicolini:2011dp}.
In conclusion, minimal length effects can be studied in a wide range of scenarios, leading to pivotal Quantum Gravity predictions that can be tested with current and future experiments.

\section*{Acknowledgements}
This work is supported by the Helmholtz International Center for FAIR within the
framework of the LOEWE program (Landesoffensive zur Entwicklung Wissenschaftlich-\"{O}konomischer
Exzellenz) launched by the State of Hesse.

\end{document}